\documentclass[12pt,times]{article}

\usepackage{times}
\usepackage{epsfig}
\usepackage{amssymb}
\usepackage{subfigure}
\usepackage{graphicx}
\usepackage{color}
\usepackage{jheppub} 

\begin{document}

\baselineskip 6mm
\renewcommand{\thefootnote}{\fnsymbol{footnote}}


\newcommand{\nc}{\newcommand}
\newcommand{\rnc}{\renewcommand}




\newcommand{\tcb}{\textcolor{blue}}
\newcommand{\tcr}{\textcolor{red}}
\newcommand{\tcg}{\textcolor{green}}


\def\be{\begin{eqnarray}}
\def\ee{\end{eqnarray}}
\def\nn{\nonumber\\}


\def\ct{\cite}
\def\la{\label}
\def\eq#1{\eqref{#1}}


\def\a{\alpha}
\def\b{\beta}
\def\g{\gamma}
\def\G{\Gamma}
\def\d{\delta}
\def\D{\Delta}
\def\e{\epsilon}
\def\et{\eta}
\def\ph{\phi}
\def\Ph{\Phi}
\def\ps{\psi}
\def\Ps{\Psi}
\def\k{\kappa}
\def\l{\lambda}
\def\L{\Lambda}
\def\m{\mu}
\def\n{\nu}
\def\th{\theta}
\def\Th{\Theta}
\def\r{\rho}
\def\s{\sigma}
\def\S{\Sigma}
\def\ta{\tau}
\def\o{\omega}
\def\O{\Omega}
\def\pr{\prime}


\def\half{\frac{1}{2}}
\def\goto{\rightarrow}

\def\na{\nabla}
\def\grad{\nabla}
\def\curl{\nabla\times}
\def\div{\nabla\cdot}
\def\pa{\partial}
\def\fr{\frac}

\def\bra{\left\langle}
\def\ket{\right\rangle}
\def\lb{\left[}
\def\lc{\left\{}
\def\ls{\left(}
\def\lp{\left.}
\def\rp{\right.}
\def\rb{\right]}
\def\rc{\right\}}
\def\rs{\right)}

\def\vac#1{\mid #1 \rangle}


\def\td#1{\tilde{#1}}
\def\check{ \maltese {\bf Check!}}


\def\Tr{{\rm Tr}\,}
\def\det{{\rm det}}
\def\text#1{{\rm #1}}


\def\bc#1{\nnindent {\bf $\bullet$ #1} \\ }
\def\ch {$<Check!>$ }
\def\ss {\vspace{1.5cm}}
\def\inf{\infty}

\begin{titlepage}

\hfill\parbox{2cm} { }

 
\vspace{1cm}

\begin{center}
{\Large \bf Correlation functions \\ of boundary and defect conformal field theories}

\vskip 1. cm
   {Chanyong Park$^{a}$\footnote{e-mail : cyong21@gist.ac.kr}}

\vskip 0.5cm

{\it $^a$ Department of Physics and Photon Science, Gwangju Institute of Science and Technology,  Gwangju  61005, Korea}

\end{center}

\vskip2cm


\centerline{\bf ABSTRACT} \vskip 4mm

Applying the holographic method, we investigate correlation functions of boundary and defect conformal field theories. To describe boundary conformal field theory, we consider an end of the world brane in an asymptotic AdS space which behaves as a boundary in the dual conformal field theory. In this holographic setup, we calculate correlation functions involving the reflection effect at the boundary. We show that, when the end of the world brane has no degrees of freedom, the holographic calculation reproduces  the correlation functions known in the boundary conformal field theory. When the boundary has nontrivial boundary entropy, we calculate one- and two-point functions nontrivially relying on the boundary entropy. We further study correlation functions of defect conformal field theory after introducing a $p$-brane. We directly derive a bulk-to-defect two-point function without introducing an image operator and determine the coefficient of the two-point function exactly in the holographic setup.

\vspace{1cm}

\vspace{2cm}

\end{titlepage}

\renewcommand{\thefootnote}{\arabic{footnote}}
\setcounter{footnote}{0}



\section{Introduction}

After the AdS/CFT correspondence or holography proposal \cite{Maldacena:1997re,Gubser:1998bc,Witten:1998qj,Witten:1998zw}, there were many attempts to figure out non-perturbative features of strongly interacting systems. The AdS/CFT correspondence claimed that a non-perturbative $d$-dimensional conformal field theory (CFT) has a one-to-one map to a classical gravity theory in a $(d+1)$-dimensional AdS space. Based on the holography, various quantum properties of strongly interacting quantum field theory (QFT) has been studied on the dual gravity side. In the present work, we apply the holographic method to boundary conformal field theory (BCFT) \ct{Cardy:1984bb,Cardy:1996xt,Cardy:2004hm,Takayanagi:2011zk,Fujita:2011fp,Jensen:2013lxa} and defect conformal field theory (DCFT) \ct{Liendo:2012hy,Gaiotto:2013nva,Antunes:2021qpy,Liendo:2019jpu,Herzog:2015ioa,Herzog:2017kkj,Herzog:2017xha,FarajiAstaneh:2021foi,Jensen:2018rxu,Chalabi:2021jud,Linardopoulos:2021rfq,Georgiou:2023yak,Nishioka:2022ook} and then investigate their correlation functions in the holographic dual gravity.

For the AdS/CFT correspondence, one of the important features is that some physical quantities of QFT can be understood by geometrical objects in the dual gravity. In the holographic setup, for example, a $q \bar{q}$-potential and entanglement entropy of strongly interacting systems have been investigated by minimal surfaces anchored at the boundary of the dual geometry \ct{Maldacena:1997re,Rey:1998ik,Lee:2009bya,Park:2009nb,Ryu:2006bv,Ryu:2006ef,Lewkowycz:2013nqa,Casini:2004bw,Casini:2006es,Park:2015hcz,Park:2015dia,Park:2018snf,Kim:2016jwu}. Similarly, it was also proposed that correlation functions of QFT are able to be described by a geodesic length connecting local operators at the boundary. Following this proposal, it was shown that geodesic lengths in the AdS space reproduce two- and three-point functions known in CFT \ct{Kim:2023fbr}. Moreover, it was also found that a geodesic length in the BTZ black hole leads to the thermal correlation functions of two-dimensional thermal CFT.

For QFT and CFT, a one-point function of a local operator usually vanishes. However, this is not the case for BCFT and DCFT. The reflection caused by a boundary and defect gives rise to a nontrivial contribution to correlation functions and leads to a non-vanishing one-point function. One- and two-point functions of BCFT were evaluated by applying the image charge method. It was shown that the BCFT's two-point function has two contributions \ct{Cardy:1996xt,Setare:2011ey}. One is a direct two-point function which is the same as that of CFT without a boundary. The direct two-point function provides a dominant contribution to BCFT's and DCFT's ones. The other is a reflected two-point function involving a subdominant reflection effect. In this case, the reflected two-point function is proportional to the square of the one-point function. In the present work, we first show that the known BCFT's correlators can be reproduced by applying the holographic method when the boundary has no degrees of freedom. Then, we study more general BCFT's correlators when the boundary has nontrivial degrees of freedom or boundary entropy.

In the holographic setup, BCFT was realized by an AdS space with an end of the world (ETW) brane which anchors to the AdS boundary \ct{Takayanagi:2011zk,Chiodaroli:2012vc,Alishahiha:2011rg,Fujita:2011fp,Nozaki:2012qd,Miyaji:2022dna,Izumi:2022opi}. Intersection between an ETW brane and the AdS boundary provides a boundary to a dual CFT. In this case, the configuration of an ETW brane in the bulk crucially relies on the ETW brane's tension which is reinterpreted as the boundary entropy on the dual BCFT point of view \ct{Calabrese:2004eu,Takayanagi:2011zk}. We show that the correlation functions known in BCFT  \ct{Cardy:1996xt} correspond to the correlators without the boundary entropy. When the boundary has a nontrivial boundary entropy, the image charge method is not applicable anymore. In the holographic setup, however, it is still possible to calculate correlators with a nonvanishing boundary entropy. In this work, after determining the ETW brane's configuration by solving the junction equation \ct{Israel:1966rt,Randall:1999ee,Randall:1999vf,Kraus:1999it,Park:2000ga,Chamblin:1999ya,Park:2020jio,Kim:2023adq}, we investigate the one- and two-point functions of the BCFT including the boundary entropy. We show that, when the boundary has a nontrivial boundary entropy, the reflected two-point function relies on the boundary entropy nontrivially. We also show that the reflected two-point function is related to the square of the one-point function with a multiplication factor depending on the boundary entropy. We also investigate thermal correlators of two-dimensional BCFT by considering an ETW brane in a three-dimensional BTZ black hole geometry.

We also investigate correlators of DCFT by applying the same holographic technique. To describe DCFT in the holographic setup, we introduce a $p$-brane in a $d$-dimensional AdS space with $p < d-1$. For $p=d-1$, a $p$-brane can be regarded as an ETW brane. At the boundary of an AdS space, a $p$-brane plays the role of a defect. Assuming that there exists an operator living in the defect, we can take into account a bulk-to-defect two-point function. In the holographic setup, it can be described by a geodesic curve connecting two operators with passing through a junction point on the $p$-brane. After evaluating the length of this geodesic curve, we calculate a bulk-to-defect two-point function in the holographic setup. We show that the holographic result without using an image charge reproduces the exactly same form of the bulk-to-defect two-point function expected by the image charge method \ct{Nishioka:2022ook}. Moreover, we determine the coefficient of bulk-to-defect two-point function exactly.

The rest of this paper is organized as follows. In Sec. 2, we discuss the holographic dual of a BCFT by introducing a ETW brane. On this geometric background, we holographically evaluate one- and two-point functions of BCFT when the boundary has no degrees of freedom. In Sec. 3, we also study the correlation functions of thermal BCFTs. In Sec. 4, we investigate BCFT's one- and two-point functions when the boundary has nontrivial boundary entropy. In Sec. 5, we further study the bulk-to-defect two-point function of DCFT. Finally, we close this work with concluding remarks in Sec. 6.



\section{Correlation functions of BCFT }

We investigate correlation functions of BCFT in the holographic setup. Before doing this, we first briefly summarize the known correlators of CFT and BCFTs. For CFTs without a boundary, a two-point function of a renormalized local operator ${\cal O}(t,\vec{w},x) $ has the following form due to the conformal symmetry 
\be
\bra {\cal O}(t_1,\vec{w}_1,x_1)  \ {\cal O} (t_2,\vec{w}_2,x_2) \ket_{CFT}  =  \fr{1}{ \ls - | t_1 - t_2 |^2 +  | \vec{w}_1 - \vec{w}_2 |^2 +|x_1 - x_2 |^2 \rs^{\D}}    ,  \la{Result:2ptfunCFT}
\ee
where $\D$ means a conformal dimension of an operator. In this case, a one-point function generally vanishes. For BCFTs, however, reflection at the boundary allows a nontrivial one-point function. Assuming that a boundary is located at  $x= x_b$, where  $x$ denotes a direction perpendicular to the boundary, the one-point function of a renormalized operator located at $\lc t, \vec{w}, x \rc$ is given by \ct{Cardy:1996xt,Cardy:2004hm}
\be
\bra  {\cal O}(t,\vec{w} ,x ) \ket_{BCFT}  \equiv \bra  {\cal O}(t,\vec{w} ,x ) | {\cal B} \ket =  \fr{1}{  (2 | x -  x_{b}  |)^{\D }} ,   \la{Result:oneptfun}
\ee
where $ \lp  | {\cal B} \ket$ indicates boundary states and $\vec{w}$ are parallel directions to the boundary. It is worth noting that the coefficient of a one-point function can have a nontrivial value for more general cases depending on the boundary state, as will be shown later. When two operators are located at $\lc t, \vec{w}_1,x_1\rc$ and $\lc t, \vec{w}_2, x_2 \rc$, a BCFT's two-point function was derived by applying the image charge method \ct{Cardy:1996xt,Cardy:2004hm,Nishioka:2022ook,Cardy:1984bb,Kastikainen:2021ybu} 
\be
&& \bra {\cal O}(t,\vec{w}_1,x_1)  \ {\cal O} (t,\vec{w}_2,x_2) \ket_{BCFT}  \nn
&& \qquad \qquad \qquad =  \fr{1}{ \ls | \vec{w}_1 - \vec{w}_2 |^2 +|x_1 - x_2 |^2 \rs^{\D}} -   \fr{1}{ \ls | \vec{w}_1 - \vec{w}_2  |^2 + | x_1 + x_2 - 2 x_b |^2 \rs^{\D}}  ,   \la{Result:twoptBCFT}
\ee
where the minus sign is introduced because an image charge has a negative charge \ct{Cardy:1996xt}. Here, the first term represents a direct two-point function, which is equivalent to a two-point function of CFT, whereas the second term appears due to reflection at the boundary.

Applying the AdS/CFT correspondence, from now on, we look into the correlation functions in the holographic setup. If we consider two local  operators, $O(t,\vec{w}_1,x_1)$ and $O(t,\vec{w}_2,x_2)$, at the $d$-dimensional boundary of a $(d+1)$-dimensional AdS space
\be
ds^2 =  \fr{R^2}{z^2}  \ls  - \ dt^2 + d \vec{w}^2 + dx^2  + d z^2  \rs .  \la{Metric:pureAdS}
\ee 
their correlation function is associated with a geodesic length connecting them. Denoting a geodesic length as $L (t,\vec{w}_1,x_1;t,\vec{w}_2,x_2)$, a two-point function is holographically determined by  \ct{Susskind:1998dq,Balasubramanian:1999zv,Louko:2000tp,Solodukhin:1998ec,DHoker:1998vkc}
\be
\bra O(t,\vec{w}_1,x_1) \ O(t,\vec{w}_2,x_2)   \ket =  e^{- \D  L (t,\vec{w}_1,x_1;t,\vec{w}_2,x_2) /R} .   \la{Proposa:Twopt}
\ee
In the AdS space in \eq{Metric:pureAdS}, a geodesic length is governed by
\be
L (t,\vec{w}_1,x_1;t,\vec{w}_2,x_2) = R \int_{x_1}^{x_2}  d x \ \fr{ \sqrt{ 1 + w'^2 + z'^2} }{z}  ,
\ee
where $w^2=\vec{w} \cdot \vec{w}$ and the prime means a derivative with respect to $x$. The minimal geodesic length satisfying the equation of motion results in the following two-point function \ct{Kim:2023fbr}
\be
\bra O (t,\vec{w}_1,x_1) \ O (t,\vec{w}_2,x_2)   \ket_{CFT}  =  \fr{\e^{2 \D} }{\ls |\vec{w}_1 - \vec{w}_2 |^2 + | x_1 - x_2|^{2} \rs^\D} .  \la{Result:2ptinCFT}
\ee
Introducing a renormalized operator defined as 
\be
{\cal O}  = \fr{O}{\e^\D}  ,			\la{Result:defreop}
\ee 
the two-point function of a renormalized operator gives rise to
\be
\bra {\cal O} (t,\vec{w}_1,x_1) \ {\cal O} (t,\vec{w}_2,x_2)   \ket_{CFT}  =  \fr{ 1 }{\ls |\vec{w}_1 - \vec{w}_2 |^2 + | x_1 - x_2|^{2} \rs^\D} ,  \la{Result:2ptinCFT}
\ee
which is equivalent to the previous CFT's result in \eq{Result:2ptfunCFT} at $t_1=t_2$.


Now, let us study correlation functions of an renormalized operator for BCFT. To apply the holographic prescription, we need to know the holographic dual of BCFTs. A $d$-dimensional BCFT can be realized by an ETW brane anchored at the boundary of a $(d+1)$-dimensional AdS space. To see more details, let us consider an ETW brane having a nontrivial tension. In this case, the energy-momentum tensor of the ETW brane is represented as
\be
T_{\m\n} = - \fr{\s  }{2 \k^2}  \g_{\m\n}  ,
\ee
where $\s$ and $\g_{\m\n}$ indicate a brane's tension and induced metric on the ETW brane. Then, the  configuration of the ETW brane is determined by a junction equation \cite{Chamblin:1999ya,Park:2020jio,Kim:2023adq}
\be
\pi^{(+)}_{\mu\nu}  - \pi^{(-)}_{\mu\nu}  = T_{\m\n} ,
\ee 
where $\pi^{(\pm)}_{\mu\nu}$ indicate a canonical momenta of $\g_{\m\n}$ on the both sides of the ETW brane. In the present setup, $\pi^{(+)}_{\mu\nu} =0$ because the outside of the ETW brane is empty. In the AdS space, the canonical momentum of $\g_{\m\n}$ reads
\be			\la{result:BoundaryStressM}
\pi^{(-)}_{\mu\nu}  = - \frac{1}{2 \k^2}  \left( K_{\mu\nu} - \g_{\mu\nu} K \fr{}{} \right) 
= - \fr{R x'}{2 \k^2 \sqrt{1 + x'^2} \ z^2},
\ee
where the prime means a derivative with respect to $z$ and an extrinsic curvature is defined as $K_{\mu\nu} = \nabla_\m n_\n$ with a unit normal vector $n_\n$.

When the ETW brane is static and we take $x$ as a function of $z$, the junction equation in the spatial sections is reduced to
\be
\fr{x'}{\sqrt{1+ x'^2  }} = R \s .    \la{Relation:junctionETW}
\ee
If the BCFT defined at $z=0$ has a boundary at $x=x_b$, the solution of the above junction equation results in
\be
x = x_b -  \fr{R \s}{\sqrt{1 - R^2 \s^2}} \  z .    \la{Result:ETWconf}
\ee
This solution determines the ETW brane's configuration as a function of $z$. Using this solution, an induced metric on the ETW brane is reduced to
\be
ds_{ETW}^2 = \fr{R^2}{z^2} \ls - dt^2 + d \vec{w}^2 +  \fr{dz^2}{1 - R^2 \s^2}\rs
= \fr{\bar{R}^2}{y^2} \ls - dt^2 + d \vec{w}^2 +  dy^2 \rs  ,
\ee
where 
\be
z = \sqrt{1 - R^2 \s^2} \ y  \quad {\rm and} \quad \bar{R} = \fr{R}{\sqrt{1 - R^2 \s^2}} .
\ee
This induced metric is equivalent to a $d$-dimensional AdS space. According to the AdS/CFT correspondence, a $(d+1)$-dimensional AdS space is the dual of a $d$-dimensional CFT and the ETW brane plays the role of a boundary of the CFT. Therefore, a $(d+1)$-dimensional AdS space with a $d$-dimensional ETW brane maps to a $d$-dimensional BCFT.

We first take into account the case of $\s=0$ for which the ETW brane extends to $t$, $z$ and $\vec{w}$ at a fixed $x=x_b$ in \eq{Result:ETWconf} and leads to a trivial boundary state, for example, a vacuum without matter. In this holographic setup, the one-point function of a renormalized operator is described by a geodesic curve connecting a local operator to a reflection point $\lc t, \vec{w}_r, x_r, z_r \rc$ on the ETW brane
\be
\bra  {\cal O}(t,\vec{w}_1 ,x_1 ) \ket_{BCFT}  = \fr{1}{\e^\D} e^{-\D L (t,\vec{w}_1,x_1;t,\vec{w}_r,x_r) /R} .   \la{Formula:One}
\ee
Since a reflection point has to be located on the ETW brane, it appears at $x_r = x_b$ for $\s=0$. At this stage, $\vec{w}_r$ and $z_r$ are not fixed yet. The geodesic curve connecting an operator at $\lc \vec{w}_1, x_1,0 \rc$ to the reflection point at $\lc \vec{w}_r,  x_b, z_r \rc$ is associated with a bulk-to-boundary Green function in the bulk. Due to the rotational symmetry in the $\vec{w}$ space, the geodesic curve does not depend on the angular coordinate of the $\vec{w}$ space. Representing the radial position of $\vec{w}$ as $w$, the geodesic length is expressed as
\be
L (t,\vec{w}_1,x_1;t,\vec{w}_r,x_b) = R \int_{x_r}^{x_1}  d x \ \fr{ \sqrt{ 1 + w'^2  +  z'^2} }{z}  , \la{Action:dualCFT}
\ee
where   $\vec{w}$ and $z$ are functions of $x$ satisfying $\vec{w}_1 = \vec{w} (x_1)$, $\vec{w}_r = \vec{w} (x_r)$ and $z_r = z (x_r)$ with $x_r =x_b$ for $\s=0$.

Since the geodesic length depends on $\vec{w}$ and $x$ implicitly, there are two conserved charges
\be
\vec{P} =  \fr{R \, \vec{w}'}{z \, \sqrt{ 1 + w'^2 + z'^2}}  \quad {\rm and} \quad H = - \fr{R}{z \, \sqrt{ 1 + w'^2 + z'^2}}  .
\ee
Introducing a turning point $z_t$ where $z'=0$ and denoting the value of $\vec{w}'$ at the turning point as $\vec{v}$, the conserved charges at the turning point are reduced to  
\be
\vec{P}  =  \fr{R \,\vec{v}}{z_t \, \sqrt{ 1 + v^2 }}  \quad {\rm and} \quad H = - \fr{R}{z_t \, \sqrt{ 1 + v^2 }} .
\la{Result:consevedH}
\ee
Comparing these conserved charges, we find that $\vec{w}'$ is given by a constant, $\vec{w}'= \vec{v}$. Since the geodesic curve has to pass through $\ls x_1, \vec{w}_1,0 \rc$ and $\lc x_b, \vec{w}_r. z_r \rc$, $\vec{v}$ is given by
\be
\vec{v} = \fr{\vec{w}_1 - \vec{w}_r}{x_1 - x_b} .
\ee
Imposing the boundary condition, $\vec{w}_1 = \vec{w} (x_1)$, $\vec{w} (x)$ is determined as a function of $x$
\be
\vec{w} (x) = \vec{w}_1 - \vec{v} \ls x_1 -x \rs  .
\ee
Moreover, $z'$ from the conserved quantities results in
\be
\fr{dz}{dx} = \pm \fr{\sqrt{1+v^2} \, \sqrt{z_t^2 - z^2}}{z}  .
\ee
Solving this equation, we can determine $z$ as a function of $x$
\be
z (x) =  \sqrt{x_1 -x} \, \sqrt{ 2  \, \sqrt{1+v^2} \, z_t  - ( 1+v^2 ) \, (x_1 -x)}  .
\ee
Imposing $z_r = z(x_b)$, the turning point $z_t$ is determined as
\be
z_t = \frac{ | x_1-x_b |^2+ | \vec{w}_1-\vec{w}_r |^2+z_r^2 }{2 \sqrt{ | x_1-x_b |^2+ | w_1-w_r |^2}}  \ .
\ee

Using the above solutions, we can rewrite the geodesic length as the integral over $z$
\be
L (t,\vec{w}_1,x_1;t,\vec{w}_r,x_b) = \left| \int_{\e}^{z_t} dz \fr{R z_t}{z \sqrt{z_t^2 - z^2}} 
+ \int_{z_r}^{z_t} dz \fr{R z_t}{z \sqrt{z_t^2 - z^2}}  \right| ,
\ee
where $\e$ is introduced as a UV cutoff. After performing this integral, we finally obtain a geodesic length connecting the boundary operator to a bulk reflection point
\be
L (t,\vec{w}_1,x_1;t,\vec{w}_r,x_b) = R \log \ls \fr{ | x_1-x_b |^2+ | \vec{w}_1- \vec{w}_r |^2+z_r^2 }{z_r \, \e} \rs .  \la{Result:GeoLeng}
\ee
Now, we require $\d L =0$ to obtain a minimal geodesic length. This fixes $\vec{w}_r$ and $z_r$ to be 
\be
\vec{w}_r = \vec{w}_1 \quad {\rm and} \quad z_r = x_1 - x_b .
\ee
Using these results, the minimal geodesic length becomes
\be
L (t,\vec{w}_1,x_1;t,\vec{w}_1,x_b)  = R \log \ls \fr{ 2 \, | x_1 - x_b | }{ \e}  \rs . \la{Result:BulkBoundary}
\ee
As a consequence, the one-point function of an renormalized operator results in
\be
\bra  {\cal O}(t,\vec{w}_1 ,x_1 ) \ket_{BCFT}  = \fr{1}{ \ls 2  | x_1 -  x_{b}  | \rs^{\D }} . \la{Result:1ptfunBCFT0}
\ee
From the holographic point of view, this result represents a bulk-to-boundary Green function and is equivalent to the BCFT's one-point function in \eq{Result:oneptfun}.

Now, let us discuss two-point functions of the BCFT. The BCFT's two-point function usually has two different contributions. The one is a direct correlation between two operators. Assuming that two operators are located at $\lc t_1, \vec{w}_1, x_1 \rc$ and $\lc t_2, \vec{w}_2, x_2 \rc$, the direct two-point function is given by \ct{Kim:2023fbr}
\be
\bra {\cal O}(t_1,\vec{w}_1,x_1)  \ {\cal O} (t_2,\vec{w}_2,x_2) \ket_{CFT}   =  \fr{1}{ \ls - |t_1 - t_2 |^2 +  | \vec{w}_1 - \vec{w}_2 |^2 +|x_1 - x_2 |^2 \rs^{\D}}  . \la{Relation:TwoptinCFT}
\ee
This gives rise to the dominant contribution to the BCFT's two-point function. The other is the correlation caused by the reflection at the boundary which is subdominant. From now on, we concentrate on the reflected two-point function. The reflected two-point function is described by a minimal geodesic curve connecting two operators passing through a reflection point on an ETW brane. Using the previous bulk-to boundary Green function, the geodesic length connecting a reflection point to two operators is given by
\be
&& L (t,\vec{w}_1,x_1;t,\vec{w}_2,x_2)   \nn
&& = R  \lb \log \ls \fr{ | \vec{w}_1 - \vec{w}_r |^2 +  | x_1 - x_r |^2 + z_r^2 }{z_r \ \e}  \rs  +  \log \ls \fr{ |  \vec{w}_2 - \vec{w}_r |^2 + |  x_2 - x_r | ^2 + z_r^2 }{z_r \ \e}  \rs  \rb ,     \la{Result:GeoLeng1}
\ee
where $x_r = x_b$ for $\s=0$. The other reflection positions, $\lc \vec{w}_r, z_r \rc$ can be determined by requiring $\d L =0$.

Varying the geodesic length in \eq{Result:GeoLeng1}, the minimal geodesic length appears when $\vec{w}_r$ and $z_r$ satisfy the following equations 
\be
0 &=& - \lb  (  w_1 - w_r )^2  + ( x_1 - x_r )^2  \rb \, \lb (w_2 - w_r )^2 + (  x_2 - x_r ) ^2  \rb + z_r^4 , \nn
0 &=& 2 \vec{w}_r  w_r^2 - 3 \left(\vec{w}_1+ \vec{w}_2\right) w_r^2+ \vec{w}_r \lb 2 x_b^2-2 \left(x_1+x_2\right) x_b+2 z_r^2+w_1^2+4 \vec{w}_1 \cdot \vec{w}_2+w_2^2+x_1^2+x_2^2\rb  \nn
&& - \vec{w}_2 \left(x_1-x_b\right){}^2-  \vec{w}_1
   \left(x_2-x_b\right){}^2 - (\vec{w}_1+ \vec{w}_2) z_r^2  - (\vec{w}_1 + \vec{w}_2) \vec{w}_1 \cdot  \vec{w}_2  .
\ee
Solving these equations, we find the reflection position at
\be
\vec{w}_r &=& \fr{\vec{w}_2 x_1 + \vec{w}_1 x_2 - (\vec{w}_1 + \vec{w}_2) x_b}{x_1 + x_2 - 2 x_b}  , \nn
z_r &=& - \fr{ \sqrt{x_1 - x_b} \, \sqrt{x_2 - x_b} \, \sqrt{(w_1 - w_2)^2 +(x_1 + x_2 - 2 x_b)^2}  }{x_1 + x_2 - 2 x_b}   .
\ee
Plugging these results into the geodesic length \eq{Result:GeoLeng1}, we finally obtain the minimal geodesic length
\be
L (t,\vec{w}_1,x_1;t,\vec{w}_2,x_2)  = R \log \ls \fr{| \vec{w}_1 - \vec{w}_2 |^2 + | x_1 + x_2 - 2 x_b |^2}{\e^2} \rs .   \la{Relation:GLBCFT}
\ee
Therefore, the resulting reflected two-point function is given by
\be
\bra {\cal O}(t,\vec{w}_1,x_1)  \ {\cal O} (t,\vec{w}_2,x_2) \ket_{reflect}  =   \fr{1}{ \ls | \vec{w}_1 - \vec{w}_2  |^2 + | x_1 + x_2 - 2 x_b |^2 \rs^{\D}}     ,
\ee
Combining the direct and reflected two-point functions, the BCFT's two-point function in the holographic setup reads for $\s=0$ 
\be
&& \bra {\cal O}(t,\vec{w}_1,x_1)  \ {\cal O} (t,\vec{w}_2,x_2) \ket_{BCFT} =  \bra {\cal O}(\vec{w}_1,x_1)  \ {\cal O} (\vec{w}_2,x_2) \ket_{CFT} - \bra {\cal O}(\vec{w}_1,x_1)  \ {\cal O} (\vec{w}_2,x_2) \ket_{reflect}  \nn
&& \qquad \qquad \qquad \qquad =  \fr{1}{ \ls | \vec{w}_1 - \vec{w}_2 |^2 +|x_1 - x_2 |^2 \rs^{\D}}
 -   \fr{1}{ \ls | \vec{w}_1 - \vec{w}_2  |^2 + | x_1 + x_2 - 2 x_b |^2 \rs^{\D}}    .
\ee
This is perfectly matched to the known BCFT's two-point function in \eq{Result:twoptBCFT}. 

Before closing this section, it is worth to noting that there is another way to rederive the reflected two-point function. In the holographic setup, we also think of an image charge. The holographic image charge method, as will be seen, is applicable only for $\s=0$. Due to the existence of the boundary at $x=x_b$, the second operator at $\lc \vec{w}_2, x_2\rc$ can have an image charge at $\lc \vec{w}_2, 2 x_b - x_2 \rc$. Then, the two-point function between the first operator and the image of the second operator becomes for CFT 
\be
\bra {\cal O}(t,\vec{w}_1,x_1)  \ {\cal O} (t,\vec{w}_2, 2 x_b - x_2) \ket_{CFT}    
=  \fr{1}{ \ls   | \vec{w}_1 - \vec{w}_2 |^2 +|x_1 + x_2 - 2 x_b  |^2 \rs^{\D}}    ,
\ee
which is equivalent to the previous reflected two-point function
\be 
\bra {\cal O}(t,\vec{w}_1,x_1)  \ {\cal O} (t,\vec{w}_2, 2 x_b - x_2) \ket_{CFT}    = \bra {\cal O}(t,\vec{w}_1,x_1)  \ {\cal O} (t,\vec{w}_2,x_2) \ket_{reflect} .
\ee 
For $\s=0$, $x_1 = x_2$ and $\vec{w}_1 = \vec{w}_2$, the reflected two-point function is the same as the square of the one-point function 
\be
 \bra {\cal O}(t,\vec{w}_1,x_1)  \ {\cal O} (t,\vec{w}_1, x_1) \ket_{reflect}  =  \left| \bra {\cal O}(t,\vec{w}_1,x_1 ) \ket_{BCFT}  \right|^2 \  .   \la{Result:12atzero}
\ee

\section{Correlation functions of two-dimensional thermal BCFT}

We take into account thermal BCFT. In the holographic setup, a two-dimensional thermal BCFT can be described by a three-dimensional BTZ black hole with an ETW brane. A Euclidean BTZ black hole's metric is given by
\be
ds^2 = \fr{R^2}{z^2} \ls -  f(z) d t^2 + dx^2 + \fr{1}{f(z)} dz^2\rs  ,
\ee
where a blackening factor is given by $f(z) = 1 - z^2/z_h^2$. In this black hole geometry, the junction equation determining the ETW brane's configuration is reduced to
\be
R \s = \fr{1}{\sqrt{z'^2 + f(z)}}  .
\ee
When the boundary is located at $x = x_b$ at $z=0$, solving the junction equation determines the ETW brane's configuration as
\be
x = x_b - z_h \log \ls \fr{\sqrt{ (1 - R^2 \s^2) z_h^2 + R^2 \s^2 z^2 } - R \s z}{z_h \sqrt{1 - R^2 \s^2}} \rs .   \la{Result:ETWFT}
\ee
For $\s=0$, the ETW brane is located at $x=x_b$. Using the obtained solution, the induced metric on the ETW brane becomes
\be
ds_{ETW}^2 = \fr{R^2}{z^2} \lb   -  \ls 1 - \fr{z^2}{z_h^2}\rs d t^2  
+ \fr{z_h^2}{ (1 - z^2/z_h^2)  \  \lc ( 1 - R^2 \s^2) z_h^2 + R^2 \s^2 z^2 \rc } dz^2 \rb ,
\ee
which is again given by a two-dimensional black hole geometry.

When the ETW brane is absent, the dual QFT of the BTZ black hole is a two-dimensional thermal CFT. For two-dimensional thermal CFT, the two-point function of a renormalized operator is  given by  \ct{Kim:2023fbr}
\be
\bra {\cal O}(t, x_1) \ {\cal O} (t, x_2)   \ket_{tCFT}  =   \lb 2 z_h \sinh \ls \fr{|x_1 - x_2|}{2 z_h}  \rs \rb^{- 2 \D} .
\ee
In the UV limit with $|x_1 - x_2| \ll z_h$, the thermal two-point function is reduced to the CFT's one in \eq{Result:2ptinCFT}. This is because finite mass and thermal effect are negligible in the UV limit. In the IR limit with $|x_1 - x_2| \gg z_h$, however, the thermal two-point function becomes
\be
\bra {\cal O}(t, x_1) \ {\cal O} (t, x_2)   \ket_{tCFT}  =   \fr{ e^{- \D  |x_1 - x_2| /  z_h } }{ z_h^{2 \D}   }  + \cdots   ,
\ee
where the ellipsis means higher-order small corrections. This shows that the thermal correlation  exponentially suppresses in the IR region. This is because the local operator in the IR region has an effective mass $m_{eff} = \D/z_h = 2 \pi \D T_H$ with Hawking temperature $T_H$, due to  the interaction with thermal fluctuations.

Now, we discuss one-point function of BCFT at finite temperature for $\s=0$. We assume that an operator with a conformal dimension $\D$ is located at $x= x_1$. Then, its one-point function is governed by
\be
L (t, x_1;t, x_b) = R \int_{x_1}^{x_b}  d x \ \fr{ \sqrt{ f  +  z'^2} }{z \sqrt{f} }  ,\la{Action:geolength}
\ee
where $x_r = x_b$ for $\s=0$. Since the geodesic length is invariant under the translation in the $x$-direction, we shift $x$ to $\bar{x} = x - x_b$ without loss of generality. Then, the geodesic length is rewritten in terms of $\bar{x}$ as
\be
L (t, x_1;t, x_b) = R \int_{x_1-x_b}^{0} d \bar{x} \ \fr{ \sqrt{ f  +  z'^2} }{z \sqrt{f} }   .
\ee
Using the conserved charge, the geodesic curve is determined by
\be
\fr{dz}{d \bar{x}} = \pm \fr{ \sqrt{f} \sqrt{z_t^2 - z^2}}{z} ,  \la{Equation:dzdx}
\ee
where we have to choose a positive sign for $\bar{x} \ge 0$ and a minus sign for $\bar{x} \le 0$. For $z_1 > z_b$ where $z' < 0$, the geodesic length can be represented as the integral over $z$ 
\be
L (t, x_1;t, x_b) = R \int_{z_r}^0  dz\fr{z_h z_t}{z \sqrt{z_h^2 - z^2} \, \sqrt{z_t^2 - z^2}} .
\ee
After applying $\bar{x} \to - \bar{x}$, we rewrite the geodesic length as
\be
 L (t, x_1;t, x_b) =  R \int_{0}^{- (x_1-x_b)} d \bar{x} \ \fr{ \sqrt{ f  +  z'^2} }{z \sqrt{f} }  =  R \int_{z_r}^0  dz\fr{z_h z_t}{z \sqrt{z_h^2 - z^2} \, \sqrt{z_t^2 - z^2}} ,
\ee
where we use  $z'>0$ in order to obtain the last result. This shows that the geodesic length is invariant under the parity transformation, $\bar{x} \to - \bar{x}$. This implies that $\bar{x}=0$ or $x=x_b$ corresponds to the turning point. As a result, the reflection point coincides with the turning point, $z_r = z_t$, for $\s=0$.

Solving \eq{Equation:dzdx}, the configuration of the geodesic curve is determined as 
\be
x (z)= c + z_h \ \tanh^{-1} \ls \fr{\sqrt{z_t^2 - z^2}}{\sqrt{z_h^2 - z^2}} \rs ,
\ee
where $c$ is an integral constant.
Requiring that the reflection point is equivalent to the turning point, the integral constant becomes $c=x_b$. Moreover, imposing an additional boundary condition, $x_1 = x(0)$, $z_t$ in terms of $x_b$ and $x_1$ is determined as 
\be
z_t  = z_h \tanh \ls \fr{x_1 -x_b}{z_h} \rs .
\ee
Plugging the obtained results into the geodesic formula, we finally obtain the minimal geodesic length
\be
L (t, x_1;t, x_r) = R \log \ls \fr{2 z_h \sinh \ls  |x_1 - x_b| /z_h \rs }{\e} \rs ,
\ee
which leads to an renormalized one-point function of thermal BCFT
\be
\bra  {\cal O}(t, x ) \ket_{tBCFT}   = \fr{1}{ (2   z_h)^\D} \fr{1}{ \sinh^\D   \ls  |x_1 - x_b| /z_h \rs } ,    \la{Result:1ptfuntBCFT}
\ee
where $z_h$ can be reinterpreted as the Hawking temperature, $T_H = 1/2 \pi z_h $.

Now, we consider a two-point function of the thermal BCFT. Assuming that two operators are located at $\lc \ta_1, x_1 \rc$ and$ \lc \ta_2, x_2 \rc$, their thermal CFT's correlation function without a boundary is given by \ct{Kim:2023fbr}
\be
\bra {\cal O}(t_1,x_1)  \ {\cal O} (t_2,x_2) \ket_{tCFT}  =  \fr{1}{(2 z_h)^{2\D} }  
\fr{1}{ | - \sinh (| t_1 - t_2 |/ 2 z_h)^{2}  +  \sinh (| x_1 - x_2 |/ 2z_h)^{2} |^\D} ,   \la{Result:Gentwopoint}
\ee
which corresponds the direct two-point function for thermal BCFT. For BCFT, there exists another subdominant correction caused by the reflection. For $\s=0$, as mentioned before, one can apply to the image charge method to evaluate a reflected two-point function. When a boundary is located at $x=x_b$, the image charge of the second operator at $\lc t_2,x_2 \rc$ appears at $\lc t_2, 2 x_b - x_2\rc$. Therefore, the reflected two-point function of BCFT is equivalent to a two-point function between the first operator and the image charge of the second operator without a boundary
\be
 \bra {\cal O}(t_1,x_1)  \ {\cal O} (t_2,x_2) \ket_{reflect} &=& \bra {\cal O}(t_1,x)  \ {\cal O} (t_2,2 x_b - x_2) \ket_{tCFT}
\ee
Using the CFT's two-point function in \eq{Result:Gentwopoint}, we find the following reflected two-point function for thermal BCFT with $\s =0$
\be
&&  \bra {\cal O}(t_1,x_1)  \ {\cal O} (t_2,x_2) \ket_{reflect} =  \nn
&& \qquad \qquad \fr{1}{(2 z_h)^{2\D} } \fr{1}{ | - \sinh^{2}  (| t_1 - t_2 |/ 2 z_h) +  \sinh ^{2} (| x_1 + x_2 - 2 x_b |/ 2 z_h) |^\D} .
\ee
As a result, the two-point function for thermal BCFT results in
\be
&& \bra {\cal O}(t_1,x_1)  \ {\cal O} (t_2,x_2) \ket_{tBCFT}    =  \bra {\cal O}(t_1,x_1)  \ {\cal O} (t_2,x_2) \ket_{tCFT} -  \bra {\cal O}(t_1,x_1)  \ {\cal O} (t_2,x_2) \ket_{reflect} \nn
&& \qquad   =   \fr{1}{(2 z_h)^{2\D} } \ls  \fr{1}{ | - \sinh^{2}  (| t_1 - t_2 |/ 2 z_h) +  \sinh ^{2} (| x_1 - x_2  |/ 2 z_h) |^\D}  \rp \nn
&& \qquad \qquad  \qquad  \qquad    \lp - \fr{1}{ | - \sinh^{2}  (| t_1 - t_2 |/ 2 z_h) +  \sinh ^{2} (| x_1 + x_2 - 2 x_b |/ 2 z_h) |^\D} \rs  ,
\ee 
For $t_1 = t_2$ and $x_1 = x_2$, the reflected correlator of the thermal BCFT is related to the one-point function 
\be
 \bra {\cal O}(t, x)  \ {\cal O} (t, x) \ket_{reflect}  =  \left| \bra {\cal O}(t, x) \ket_{tBCFT}  \right|^2 \  .   \la{Result:12atfinite}
\ee
For $\s=0$, consequently, the reflected two-point function is given by the square of the one-point function both at zero temperature in \eq{Result:12atzero} and finite temperature in \eq{Result:12atfinite}.

\section{BCFT's correlators with boundary entropy}

Now, we take into account BCFT with a boundary having nontrivial degrees of freedom, $\s \ne0$. In this case, the boundary can have nontrivial entanglement entropy which was known as the boundary entropy. For $\s \ne 0$, a reflection point is not the same as a turning point. From the ETW brane's configuration in \eq{Result:ETWconf}, the reflection point $x_r$ in the $x$-direction is given by a function of $z_r$
\be
x_r = x_b -  \fr{R \s}{\sqrt{1 - R^2 \s^2}} \  z_r .   \la{Result:funxrzr}
\ee
In addition, the bulk-to-boundary geodesic length in \eq{Result:GeoLeng} is generalized into
\be
L (t,\vec{w}_1,x_1;t,\vec{w}_r,x_r) = R \log \ls \fr{ | x_1-x_r |^2+ | \vec{w}_1-\vec{w}_r |^2+z_r^2 }{z_r \, \e} \rs .    \la{Result:gGeoLeng}
\ee
After substituting \eq{Result:funxrzr} into \eq{Result:gGeoLeng}, variation of the geodesic length with respect to $\vec{w}_r$ and $z_r$ determines the reflection point as
\be
\vec{w}_r = \vec{w}_1 \quad {\rm and} \quad z_r =    \sqrt{1 - R^2 \s^2} \  (x_1 - x_b) .
\ee
As a result, the BCFT's one-point function for $\s \ne 0$ results in
\be
\bra  {\cal O}(t,\vec{w} ,x_1 ) \ket_{BCFT}  = \fr{A^\D}{ \ls 2  | x_1 -  x_{b}  | \rs^{\D }},    \la{Result:1ptfunBCFT}
\ee
where a coefficient $A$ is given by
\be
A =  \sqrt{ \fr{1 - R \s }{1  + R \s } }   .
\ee

In Ref. \ct{Herzog:2017xha}, it was shown that $A^\D$ plays a role in the bulk conformal block decomposition of the two-point function. Another implication comes from the entanglement entropy of BCFT. For a two-dimensional BCFT, the entanglement entropy is given by \ct{Calabrese:2004eu,Azeyanagi:2007qj,Fujita:2011fp,Affleck:1991tk}
\be
S_E =  \fr{c}{6}  \log \fr{l}{\e}+ 2 g  ,
\ee
where $l$ and $g$ correspond to a subsystem size and boundary entropy, respectively. When we take into account a subsystem in the region $x_b \le x \le x_1$ at $z=0$,  the entanglement entropy in the present holographic setup is given by
\be
S_E = \fr{L (t,x_1;t,x_r) }{4 G}  = \fr{c}{6}  \log \ls \fr{x_1 - x_b}{\e} \rs +  \fr{c}{6}   \log \ls \fr{2}{A} \rs   ,
\ee
where a central charge is given by $c = 3 R/2 G$. In this case, the boundary entropy $g$ is connected to the ETW brane's tension $\s$
\be
g = \fr{c}{12} \log 2 + \fr{c \, \D}{24} \log \ls \fr{1+R \s }{1 - R \s} \rs  .
\ee
As a result, the brane's tension $\s$ can be associated with the entropy of the boundary states.

We further take into account a two-point function of BCFT. For simple calculation, we assume that two operators are located at the same $x$ position, $x_1 = x_2$, with $\vec{w}_1 \ne \vec{w}_2$. 
After plugging \eq{Result:funxrzr} into the geodesic length \eq{Result:GeoLeng1} connecting the reflection point to two local operators and varying it with respect to $z_r$ and $\vec{w}_r$, we find that the geodesic length has a minimum at
\be
\vec{w}_r = \fr{\vec{w}_1 + \vec{w}_2}{2} \quad {\rm and} \quad
z_r =  \fr{\sqrt{1-R^2 \s^2}}{2}  \sqrt{ | \vec{w}_1 - \vec{w}_2 |^2 + 4 | x_1 - x_b |^2}  .
\ee
Utilizing these results, the reflected two-point function for $\s \ne 0$ is reduced to
\be
\bra {\cal O}(t,\vec{w}_1,x_1)  \ {\cal O} (t,\vec{w}_2,x_1) \ket_{reflect}   =   \fr{ (1 - R^2 \s^2) ^{\D} }{ \ls \sqrt{| \vec{w}_1 - \vec{w}_2  |^2 +4  | x_1  -  x_b |^2} + 2 R \s | x_1 - x_b | \rs^{2 \D}}    .
\ee
As a consequence, the BCFT's two-point function with the nontrivial boundary entropy becomes
\be
&& \bra {\cal O}(t,\vec{w}_1,x_1)  \ {\cal O} (t,\vec{w}_2,x_1) \ket_{BCFT}  \nn
&& \qquad \qquad \qquad  =  \fr{1}{ | \vec{w}_1 - \vec{w}_2 | ^{2 \D}} -   \fr{(1 - R^2 \s^2) ^{\D } }{ \ls \sqrt{| \vec{w}_1 - \vec{w}_2  |^2 +4  | x_1  -  x_b |^2} + 2 R \s | x_1 - x_b | \rs^{2 \D}}    .   \la{Result:BCFTfunsne0}
\ee
This result shows that, when the boundary has a nontrivial boundary entropy, the BCFT's two-point function \eq{Result:twoptBCFT} for $x_1=x_2$ is generalized into \eq{Result:BCFTfunsne0}. In this case, the holographic result for the reflected two-point function cannot be explained by the naive image charge method. When $\vec{w}_1 = \vec{w}_2$, the reflected two-point function for $\s \ne 0$ is again given by the square of the one-point function 
\be
 \bra {\cal O}(t,\vec{w},x_1)  \ {\cal O} (t,\vec{w}, x_1) \ket_{\rm reflect}  = 
 \left| \bra {\cal O}(t,\vec{w},x_1 ) \ket_{\rm BCFT}  \right|^2 \  .  \la{Result:MF}
\ee

\section{Correlation functions of DCFT}

For DCFT, one- and two-point functions have been studied  \ct{Liendo:2012hy,Gaiotto:2013nva,Herzog:2015ioa,Herzog:2017kkj,Herzog:2017xha,Nishioka:2022ook}. Assuming that the defect is located at $\vec{x}=0$ and denoting perpendicular directions to the defect as $\vec{w}$, a one-point function and bulk-to-defect two-point function are given by 
\be
\bra {\cal O}^{\D} (\vec{x},\vec{w})  \ket_{DCFT}  &=& \fr{a_{\D}}{ |\vec{x}|^{\D}}   ,  \la{Result:oneptdCFT}  \\
\bra {\cal O}^{\D} (\vec{x},\vec{w}) \  \bar{{\cal O}}^{\D_d} (0,\vec{w}_d)  \ket_{DCFT}  &=&  \fr{b_{\D,\D_d}}{ |\vec{x}|^{\D-\D_d}  \  \ls  | \vec{w} - \vec{w}_d |^2 +  | \vec{x} |^2\rs^{\D_d}} ,   
\ee
where ${\cal O}^{\D} (\vec{x},\vec{w})$ is an bulk operator with a conformal dimension $\D$ and $ \bar{{\cal O}}^{\D_d} (0,\vec{w}_d) $ implies an operator living in the defect with a conformal dimesion $\D_d$. The bulk-to-defect two-point function was obtained by calculating a bulk-bulk-defect three-point function \ct{Nishioka:2022ook}. If one splits a bulk operator ${\cal O}^\D (\vec{x},\vec{w})$ into ${\cal O}^{\D/2} (\vec{x},\vec{w})$ and ${\cal O}^{\D/2} (-\vec{x},\vec{w})$, where ${\cal O}^{\D/2} (- \vec{x},\vec{w} )$ corresponds to the image of ${\cal O}^{\D/2} (\vec{x},\vec{w})$, a bulk-bulk-defect three point function is reduced to
\be
\bra {\cal O}^{\D/2} (\vec{x},\vec{w}) \ {\cal O}^{\D/2} (-\vec{x},\vec{w}) \ {\cal O}_d^{\D_d} (0,\vec{w}_d) \ket_{CFT} =  \fr{c_{\D/2,\D/2,\D_d}}{ 2^{\D-\D_d}  \  |\vec{x}|^{\D-\D_d}  \   \ls  |\vec{w} - \vec{w}_d |^2 +  |\vec{x} |^2\rs^{\D_d}} .
\ee
This three-point function is the same as the bulk-to-defect two-point function up to overall multiplication factor. In this section, we rederive the bulk-to-defect two-point function using the holographic method without introducing an image operator. 

To describe DCFT in the holographic setup, we take into account a $p$-brane with $p < d-1$ to describe a $(p-1)$-dimensional defect at the $d$-dimensional boundary space. Assuming that a $p$-brane extends $t$, $z$ and $(p-1)$ $\vec{w}$-directions, $\vec{x}$ and $\vec{w}$ indicate perpendicular and parallel directions to the $p$-brane. When a $p$-brane is located at $\vec{x}=0$, the dual theory becomes CFT with a defect at $\vec{x}=0$. In this holographic setup, a one-point function of DCFT can be easily evaluated by calculating a geodesic length between the bulk operator and a reflection point living in the $p$-brane. Assuming that an operator $ {\cal O} (t, \vec{x}, \vec{w} )$ is located at $\lc \vec{x} , \vec{w} , z \rc=\lc \vec{x} , \vec{w} , 0  \rc$, we can expect that a reflection point is located at $\lc \vec{0} , \vec{w} , z_r  \rc$ to have a shortest geodesic length. This is because a reflection point freely moves on the $p$-brane. As a result, a one-point function of DCFT is described by a geodesic curve connecting an operator at $\lc \vec{x} , \vec{w} , 0  \rc$ to a reflection point at $\lc \vec{0} , \vec{w} , z_r  \rc$, where $z_r$ is not fixed yet. This is the same as the previous one-point function of a boundary CFT in \eq{Result:1ptfunBCFT0}. Setting $x_b = 0$ in \eq{Result:1ptfunBCFT0}, the DCFT's one-point function results in  
\be
\bra  {\cal O}(t, \vec{x}, \vec{w} ) \ket_{DCFT}  = \fr{a_{\D}  }{ |\vec{x}  |^\D   }  ,
\ee
with the following coefficient
\be
 a_{\D} =  \fr{1}{2^{\D} } ,
\ee
where $\D$ is a conformal dimension of the operator.

In the holographic setup, a bulk-to-defect two-point function is determined by a minimal geodesic passing through a junction point.  Denoting the position of a junction point as $\lc \vec{x}= \vec{0}, \vec{w}_r, z_r \rc$, a bulk-to-defect two-point function is determined by the sum of two geodesics
\be
L  (\vec{x},\vec{w} ; 0,\vec{w}_d)= R \lb \D \, \log \ls \fr{|\vec{x}|^2 + |\vec{w} - \vec{w}_r|^2 + z_r^2| }{z_r \, \e} \rs +  \D_d \, \log \ls \fr{ |\vec{w}_d - \vec{w}_r|^2 + z_r^2| }{z_r \, \e} \rs \rb ,  \la{Result:geodCFT}
\ee
where a geodesic length is determined by \eq{Result:GeoLeng}. In this case, the first term describes a geodesic from the bulk operator to a junction point, while the other indicates the geodesic from a defect operator to a junction point. At this stage, the position of a junction point is not fixed yet. To determine a junction point, we vary $L$ and find a junction point satisfying $\d L=0$. After some calculation, we find
\be
\vec{w}_r &=& \frac{ (\Delta -\D_d) \  \ls | \vec{w} - \vec{w}_d |^2 + | \vec{x} |^2  \rs   \ \vec{w}   + 2 \D_d  \ | \vec{x}|^2  \ \vec{w}_d }{ (\Delta -\D_d) \ |\vec{w} - \vec{w}_d |^2 + (\D_d+\Delta ) \ | \vec{x} |^2   }     , \\
z_r &=& \frac{  \sqrt{\Delta^2 -\D_d^2}  \  \left( | \vec{w}- \vec{w}_d |^2+ | \vec{x} |^2\right) \ | \vec{x}| }{  (\Delta -\D_d) \  |\vec{w} - \vec{w}_d |^2 +  (\D_d+\Delta ) \  | \vec{x} |^2 } .
\ee
Substituting these values into \eq{Result:geodCFT}, we finally obtain a renormalized bulk-to-defect two-point function
\be
\bra {\cal O}^{\D} (\vec{x},\vec{w}) \  \bar{{\cal O}}^{\D_d} (0,\vec{w}_d)  \ket_{DCFT}    &=&  e^{- L (\vec{x},\vec{w} ; 0,\vec{w}_d) /R}  \nn
&=&  \fr{b_{\D,\D_d}}{ |\vec{x}|^{\D-\D_d}  \ \ls  | \vec{w} - \vec{w}_d |^2 +  | \vec{x} |^2\rs^{\D_d}}  ,
\ee
with the following coefficient
\be
b_{\D,\D_d} = \fr{ \sqrt{ (\D^2 - \D_d^2 )^\D  } }{2^\D \ \D^\D}  \sqrt{ \ls \fr{  \D + \D_d  }{  \D - \D_d  } \rs^{\D_d}  }    .   \la{Result:codefect}
\ee
This holographic result for the bulk-to-defect two-point function is perfectly matched to that in Ref. \ct{Nishioka:2022ook}. Moreover, the holographic calculation exactly determines the coefficient of the two-point function in \eq{Result:codefect}.


\section{Discussion}

For CFT without a boundary, the one-point function of a local operator generally vanishes. For BCFT and DCFT, however, the reflection caused by the boundary or defect gives rise to nontrivial contribution to the correlation functions. For BCFT and DCFT, therefore, the reflection provides nontrivial contribution to one- and two-point functions. In this work, we have investigated various correlation functions of the BCFT and DCFT by applying the holographic method.  

To describe $d$-dimensional BCFT in the holographic setup, we considered a $(d+1)$-dimensional AdS space with a $d$-dimensional ETW brane, which plays a role of the boundary of dual CFT. In this case, the configuration of an ETW brane was determined by solving the junction equation which crucially depends on the tension of the ETW brane. In this holographic setup, the BCFT's one-point function  is represented as a minimal geodesic curve reflected by the ETW brane. We showed that, when the ETW brane has a zero tension, the holographic method reproduces the exactly same one-point function known in the BCFT \ct{Cardy:1996xt}. We further investigated the BCFT's correlation functions when the boundary has a nontrivial tension. After showing that the non-vanishing ETW brane's tension is associated with the boundary entropy, we calculated the one-point function of BCFT with a nontrivial boundary entropy. We show that the boundary entropy modifies the coefficient of the BCFT's one-point function.

We also studied the BCFT's two-point functions. For BCFT, the two-point function has two contributions. The leading contribution comes from the CFT's two-point function without the boundary. The other is the reflected two-point function caused by the reflection and then provides the subdominant correction. When the boundary has no degrees of freedom, we showed that we can apply the image charge method to the holographic setup, similar to that of BCFT. As a result, the reflected two-point function is equivalent to the two-point function between one operator and the image of the other operator. In this case, we showed that the reflected two-point function is equivalent to the square of the one-point function.  We also studied the thermal BCFT's one- and two-point functions at finite temperature.

We further investigated the BCFT's correlation functions when the boundary has nontrivial boundary entropy. In this case, we showed that the image charge method does not work anymore. Nevertheless, we calculated the BCFT's one- and two-point functions by finding a reflection point in the holographic setup. We showed that, when the boundary has nontrivial boundary entropy, the reflected two-point function is significantly modified. When two operators are overlapped, we also showed that the reflected two-point function is given by the square of the one-point function with a nontrivial multiplication factor crucially depending on the boundary entropy.

By applying the same holographic method to DCFT, lastly, we further investigated the bulk-to-defect two-point function for DCFT. To describe DCFT in the dual gravity, we introduced a $p$-brane which plays a defect on the dual QFT side. Assuming a defect operator living in the defect, we can take into account a bulk-to-defect two-point function. In the holographic setup, it can be described by a geodesic curve connecting two operators with passing through a junction point on the $p$-brane. We found a bulk-to-defect two-point function by evaluating the geodesic length directly without introducing an image charge. We showed that the holographic result reproduces the exactly same form as that derived by the image charge method \ct{Nishioka:2022ook}. We further determined the coefficient of two-point function exactly.

\vspace{0.5cm}

{\bf Acknowledgement}

C. P. was supported by the National Research Foundation of Korea(NRF) grant funded by the Korean government (No. NRF-2019R1A2C1006639).



%

\end{document}